\documentclass[fleqn,usenatbib]{mnras}

\usepackage{newtxtext,newtxmath}

\usepackage[T1]{fontenc}

\DeclareRobustCommand{\VAN}[3]{#2}
\let\VANthebibliography\thebibliography
\def\thebibliography{\DeclareRobustCommand{\VAN}[3]{##3}\VANthebibliography}

\usepackage{orcidlink}
\usepackage{graphicx}	
\usepackage{amsmath}	
\usepackage{amsfonts}
\usepackage{bm}
\usepackage{multirow} 
\usepackage{ulem}
\usepackage{comment}

\newcommand{\hmpc}{\,h^{-1}\,\mathrm{Mpc}}
\newcommand{\hgpc}{\,h^{-1}\,\mathrm{Gpc}}
\newcommand{\ihmpc}{\,h\,\mathrm{Mpc}^{-1}}
\newcommand{\lsub}{L_\mathrm{sub}}
\newcommand{\lcell}{l^\mathrm{cell}}
\newcommand{\ngrid}{N_\mathrm{grid}}
\newcommand{\nrebin}{N_\mathrm{rebin}}
\newcommand{\nbatch}{n_\mathrm{batch}}
\newcommand{\dini}{\delta_\mathrm{target}(z=10)}
\newcommand{\dfin}{\delta_\mathrm{input}(z=0)}
\newcommand{\drec}{\delta_\mathrm{recon}}

\title[Optimal scales for CNN reconstruction]{Standard Reconstruction Shifts the Optimal Input Scale for CNN-Based Density-Field Reconstruction}

\author[K. Nakashima et al.]{
Koichiro Nakashima$^{1}$\orcidlink{0009-0004-7367-3772}\thanks{E-mail: nakashimak@ccs.tsukuba.ac.jp},
Kiyotomo Ichiki$^{2,3,4}$\orcidlink{0000-0003-1365-8568} \thanks{Email: ichiki.kiyotomo.a9@f.mail.nagoya-u.ac.jp},
Atsushi J. Nishizawa$^{5,3}$\orcidlink{0000-0002-6109-2397}\thanks{Email: atsushi.nishizawa@iar.nagoya-u.ac.jp}, 
\\
$^1$ Center for Computational Sciences, University of Tsukuba,
Ten-nodai, 1-1-1, Tsukuba, Ibaraki 305-8577, Japan,\\
$^2$ Department of Physics, Nagoya University, Furocho, Chikusa, Nagoya, Aichi 464-8602, Japan,\\
$^3$ Kobayashi Maskawa Institute, Nagoya University, Furocho, Chikusa, Nagoya, Aichi, 464-8602, Japan\\
$^4$ Institute for Advanced Research, Nagoya University, Furocho, Chikusa, Nagoya, Aichi, 464-8602, Japan\\
$^5$ DX Center, Gifu Shotoku Gakuen University, Takakuwanishi, Yanaizu, Gifu, 501-6194, Japan\\
}

\date{Accepted XXX. Received YYY; in original form ZZZ}

\pubyear{\the\year{}}

\begin{document}
\label{firstpage}
\pagerange{\pageref{firstpage}--\pageref{lastpage}}
\maketitle

\begin{abstract}
We investigate convolutional neural network (CNN) methods for reconstructing the high-redshift density field from late-time large-scale structure, focusing on how the physical scale of the CNN input changes when standard first-order reconstruction is applied beforehand. Using dark-matter-only $N$-body simulations, we compare three approaches: a single-input CNN, a dual-input CNN combining two physical scales, and a single-input CNN applied to the density field after standard reconstruction. We vary the physical side length of the input sub-box over $\lsub\sim38$--$380~h^{-1}\mathrm{Mpc}$ while keeping its numerical size fixed at $39^3$ voxels, allowing us to examine the trade-off between spatial context and resolution. For the CNN applied directly to the evolved density field, the reconstruction performs best at $\lsub\sim150$--$200~h^{-1}\mathrm{Mpc}$. After standard reconstruction, however, the preferred scale shifts to $\lsub\sim38$--$114~h^{-1}\mathrm{Mpc}$. The single-input CNN after standard reconstruction consistently outperforms both the single- and dual-input CNNs without standard reconstruction according to the normalized loss, density probability distribution, Kullback--Leibler divergence, residual field, and Fourier-space correlation. These results indicate that coherent large-scale displacements are more efficiently recovered by perturbative reconstruction, while the CNN is better suited to modelling the remaining quasi-linear and non-linear evolution on smaller scales. The preferred post-reconstruction input range includes the effective receptive scale of approximately $60~h^{-1}\mathrm{Mpc}$ adopted in previous hybrid reconstruction studies. Our findings therefore support a physically motivated separation of scales between analytic and data-driven reconstruction and demonstrate the advantage of combining the two approaches.
\end{abstract}



\section{Introduction}
The large-scale structure (LSS) observed today formed from nearly uniform initial conditions via gravitational instability. This evolution encodes information on dark matter, dark energy, and baryons, motivating efforts to recover the initial density field from late-time data for improved early-Universe constraints and phase-level comparisons.

If gravitational evolution can be approximately inverted, one obtains an estimate of the linear density field prior to non-linear collapse. A key example is the baryon acoustic oscillation (BAO) feature, a relic of sound waves in the photon–baryon fluid that provides a standard ruler for cosmic expansion \citep{Weinberg2013}. At late times, bulk flows and mode coupling smear the BAO signal; reconstruction sharpens the feature and improves distance measurements. More generally, access to the initial field enhances probes of primordial non-Gaussianity \citep{Bartolo2004} and related physics.

A further motivation, central to this work, is to enable \textit{direct} cross-epoch tests by comparing reconstructed linear fields from LSS with independent high-redshift estimates. The Kamionkowski--Loeb method \citep{Kamionkowski1997} infers large-scale primordial modes from CMB polarization induced by Thomson scattering in clusters \citep[e.g.][]{Seto2000,Portsmouth2004,2022PhRvD.105f3507I}. Combined with LSS reconstruction, this enables phase-matched comparisons that can, in principle, overcome cosmic-variance limitations.

While forward gravitational evolution is well understood, the inverse problem is challenging due to non-linearities, shell crossing, and the long-range nature of gravity. A range of approaches has been developed, including perturbative, iterative, and forward-modeling methods. Early work by \citet{Nusser1992} used the Zel’dovich approximation \citep{Zel'dovich1970} for backward evolution. A major advance was the BAO reconstruction method of \citet{Eisenstein2007}, based on the linear perturbation theory (often referred to as \textit{standard reconstruction}), which has since been widely studied (e.g. \citet{Seo2008,Noh2009,Padmanabhan2009,Schmittfull2015}) and applied observationally (e.g. \citet{Padmanabhan2009,Xu2013,Hinton2017}).

Subsequent developments include non-linear reconstruction schemes based on displacement-field recovery. \citet{Zhu2016} demonstrated in 1D that suppressed linear modes can be restored, later extended to 3D with multigrid methods \citep{Zhu2017}. Further studies showed improved BAO recovery, information gain, and robustness to realistic tracers and biases (e.g. \citet{Pan2017,Wang2017,Yu2017,WangPen2019,Zhu2018}). Alternative approaches include iterative schemes \citep{Schmittfull2017}, Bayesian reconstruction \citep{Seljak2017}, and methods addressing non-linear mapping and redshift-space effects \citep{Shi2018,Wang2020}.

Machine-learning techniques have also been incorporated into forward-modelling and field-level inference frameworks. \citet{2018JCAP...10..028M} combined neural-network modelling of the galaxy–matter connection with differentiable forward modelling to reconstruct initial density fields, while CosmicRIM used recurrent inference machines together with differentiable cosmological simulations to accelerate early-Universe reconstruction \cite{2021arXiv210412864M}. In parallel, Bayesian field-level approaches, such as the BORG framework, provide a probabilistic route to reconstructing initial conditions from non-linearly evolved density fields \cite{10.1093/mnras/stt449}.

More recently, machine learning approaches, particularly convolutional neural networks (CNNs; \citealt{Krizhevsky2012}), have been applied to this problem. \citet{Mao2021} developed a CNN that infers initial density fields from late-time sub-volumes, recovering large-scale modes and enhancing the BAO signal. Expanding on this idea, \citet{Shallue2023} pointed out a fundamental tension between the long-range nature of gravity 
and the limited receptive fields of CNNs. 
Rather than simply increasing receptive fields, at the cost of computation or resolution, they introduced a \textit{hybrid} strategy: 
first applying standard first-order reconstruction to reverse large-scale bulk flows, 
then using a CNN to learn residual non-linear corrections. 
This two-step framework significantly improved reconstruction accuracy, 
particularly at smaller scales (\(k \sim 0.4~h~\mathrm{Mpc}^{-1}\)).

\citet{Chen2023} extended this hybrid approach, systematically studying the dependence of CNN performance on reconstruction parameters and shot noise. Using reconstructed density fields from $N$-body simulations, they trained an eight-layer CNN that recovers initial conditions with high accuracy up to \(k \sim 0.5~h~\mathrm{Mpc}^{-1}\), while improving redshift-space corrections and maintaining robustness across cosmologies, albeit with degradation at high shot noise. \citet{Parker2025} further proposed a hybrid model combining standard BAO reconstruction with CNNs trained on sub-grids using mock halo and galaxy catalogues. 
Very recently, \citet{2026arXiv260315732B} compared traditional Zel’dovich reconstruction, explicit field-level inference based on differentiable forward modelling, and implicit field-level inference using a CNN correction to standard reconstruction in a field-level BAO analysis. Their results further support the effectiveness of combining perturbative reconstruction with machine-learning corrections in realistic galaxy samples.

In our previous work, \citet{Nakashima2025MNRAS}, we systematically investigated the dependence of reconstruction performance on the spatial scale of the CNN input. Keeping the number of grid points fixed, we found that $\sim150$–$200\hmpc$ is the optimal input scale. We also introduced a framework that enables CNNs to incorporate multi-scale information without relying on hybrid methods such as those of \citet{Shallue2023} and \citet{Parker2025}. By combining a high-resolution, small-scale input with a coarse, large-scale input, the network significantly improves reconstruction accuracy—particularly on small scales—relative to the best single-input model, despite using the same parent simulation. We refer to this approach as the \textit{dual-input} CNN. These results highlight the importance of input scale and network design in reconstruction tasks.

In this work, we revisit the dependence on input scale after applying standard reconstruction. We address two main questions: (i) how the optimal input scale shifts after standard reconstruction, and (ii) whether the Zel’dovich approximation or CNN-based methods are more effective for capturing large-scale information. To this end, we compare multiple metrics across different input scales, with and without standard reconstruction, as well as for the dual-input CNN. We emphasize that our objective is not to provide a comprehensive comparison with the hybrid methods of \citet{Shallue2023} and \citet{Parker2025}, nor to quantify the ultimate level of BAO recovery. Instead, this study is designed to isolate the impact of standard reconstruction on the optimal CNN input scale and to investigate how large-scale information is captured by different reconstruction strategies.

The structure of this paper is as follows. In Section~\ref{sec:methods}, we describe the methodology, including the dataset, preprocessing, CNN architecture, and training procedure. Section~\ref{sec:results} presents the main results, focusing on the dependence of reconstruction accuracy on input scale across multiple statistics. Section~\ref{sec:discussions} discusses the implications and limitations of our approach. Finally, Section~\ref{sec:conclusions} summarizes our findings and outlines future directions.


\section{Methods}
\label{sec:methods}

Our goal is to reconstruct the high-redshift dark matter density field at $z=10$ from the evolved density field at $z=0$. For each target grid point, a convolutional neural network (CNN) receives a cubic region centred on that point and predicts the corresponding density at $z=10$. We compare three approaches: a single-input CNN applied directly to the $z=0$ density field, a dual-input CNN that combines two physical scales, and a hybrid approach in which standard reconstruction is first applied to the $z=0$ field and the reconstructed field is then provided to the CNN. By varying the physical size of the CNN input while keeping its numerical dimensions fixed at $39^3$ voxels, we examine which spatial scales contain the information required for each approach.

\subsection{Standard reconstruction}
\label{subsec:stdrec}
We define the matter density contrast as
\begin{align}
\delta(\bm{x},t)
\equiv
\frac{\rho(\bm{x},t)-\bar{\rho}(t)}{\bar{\rho}(t)},
\end{align}
where $\rho(\bm{x},t)$ is the matter density at Eulerian position $\bm{x}$ and $\bar{\rho}(t)$ is its cosmic mean.

In Lagrangian perturbation theory (LPT), a matter element initially located at the Lagrangian coordinate $\bm{q}$ moves to the Eulerian position
\begin{align}
\bm{x}(\bm{q},t)=\bm{q}+\bm{\Psi}(\bm{q},t),
\end{align}
where $\bm{\Psi}$ is the displacement field. At first order, the displacement is irrotational and obeys
\begin{align}
\bm{\nabla}_{\bm{q}}\cdot\bm{\Psi}^{(1)}(\bm{q},t)
=-\delta_{\rm L}(\bm{q},t),
\label{eq}
\end{align}
where $\delta_{\rm L}$ is the linear density contrast. This relation defines the Zel'dovich approximation \citep{Zel'dovich1970}. In Fourier space, it becomes
\begin{align}
\bm{\Psi}^{(1)}(\bm{k},t)
=\frac{i\bm{k}}{k^2}\delta_{\rm L}(\bm{k},t).
\end{align}
The factor $k^{-2}$ shows that the displacement at a given position depends non-locally on the surrounding density field and is particularly sensitive to long-wavelength modes.

Following \citet{Shallue2023}, the variance of the Zel’dovich displacement is given by
\begin{align}
\int \frac{\mathrm{d}k}{2\pi^2}P_{\rm L}(k,t),
\end{align}
where $P_{\rm L}(k,t)$ is the linear matter power spectrum. In a $\Lambda$CDM cosmology, the dominant contribution arises near $k\sim0.01\hmpc$, corresponding to a wavelength $2\pi/k\sim600\hmpc$. Large-scale bulk displacements are therefore sourced by modes extending over several hundred $h^{-1}\mathrm{Mpc}$. Such information is naturally included in methods based on the Zel'dovich approximation but is difficult for a CNN to infer when its input is restricted to a finite local region.

Standard reconstruction \citep{Eisenstein2007,2018MNRAS.478.1866H} exploits this property to partially reverse non-linear evolution and sharpen the baryon acoustic oscillation (BAO) feature. The method estimates the large-scale displacement field from a smoothed density field and shifts tracers accordingly, thereby undoing bulk flows and reducing mode coupling (see also \citet{Schmittfull2015}).

In this work, we perform standard reconstruction using \texttt{nbodykit} \citep{Hand2018AJ} with the \textit{Lagrangian Growth-Shift} (LGS) scheme. The reconstructed density field is defined as
\begin{align}
\delta_\mathrm{rec}(\bm{x}) \equiv \delta_\mathrm{disp}(\bm{x}) - \delta_\mathrm{rand}(\bm{x}),
\end{align}
where $\delta_\mathrm{disp}$ is obtained by displacing the clustered field by the negative Zel’dovich displacement, and $\delta_\mathrm{rand}$ is constructed by applying the same displacement to a uniform random field. This subtraction removes the uniform component and isolates the reconstructed fluctuations. The resulting reconstructed field $\delta_\mathrm{rec}$ is then used as input to the CNN in our hybrid framework.
The CNN is still trained to predict the density at $z=10$ directly, but it only needs to learn the corrections that remain after standard reconstruction has removed part of the large-scale evolution.

\subsection{Convolutional Neural Network}
\label{subsec:cnn}
A neural network (e.g. \citet{LeCun2015,Goodfellow2016}) is a parametric function that maps inputs to outputs through layers of neurons. Each layer $i\in\mathbb{N}$ applies a linear transformation from an input vector $\bm{x}_{i-1} \in\mathbb{R}^n$ to an output vector $\bm{x}_{i} \in\mathbb{R}^m$
followed by a non-linear activation:
\begin{align} \bm{x}_i=f(\bm{W}_i\bm{x}_{i-1}+\bm{b}_i), 
\end{align}
where $\bm{W}_i \in\mathbb{R}^{m\times n}$ and $\bm{b}_i\in\mathbb{R}^m $ are trainable parameters, and $f:\mathbb{R}^m\to\mathbb{R}^m$ is a non-linear function such as ReLU \citep{Nair2010}. Stacking layers enables the model to learn hierarchical representations. Training optimizes these parameters using input–output pairs.

Convolutional neural networks (CNNs; e.g. \citet{Krizhevsky2012}) replace dense operations with convolutions. The $l$-th channel in layer $i$ is
\begin{align} \bm{x}^l_i=f\left(\sum_k{\bm{W}^{l,k}_i\otimes\bm{x}^k_{i-1}+\bm{b}_i^l}\right), 
\end{align}
where $\otimes$ denotes convolution. This localized operation efficiently captures spatial correlations, making CNNs well suited to cosmological density fields. Pooling layers are often used to reduce resolution and suppress small-scale noise.

\begin{table}
 \centering
\begin{tabular}{lcccc}
\hline 
Layer & Kernel size & Output shape & Stride \\
\hline \hline
input & None & ($\nbatch$, 39, 39, 39, 1) & None  \\\hline
conv1 & (3, 3, 3) &($\nbatch$, 20, 20, 20, 32)& (2, 2, 2) \\ \hline
conv2 & (3, 3, 3) &($\nbatch$, 20, 20, 20, 32) & (1, 1, 1)   \\\hline
conv3 & (3, 3, 3) &($\nbatch$, 10, 10, 10, 64) & (2, 2, 2)    \\\hline
conv4 & (3, 3, 3) &($\nbatch$, 10, 10, 10, 64) & (1, 1, 1)  \\\hline
conv5 & (3, 3, 3) &($\nbatch$, 5, 5, 5, 128) & (2, 2, 2) \\\hline
conv6 & (3, 3, 3) &($\nbatch$, 5, 5, 5, 128)  & (1, 1, 1)  \\\hline
conv7 & (1, 1, 1) &($\nbatch$, 5, 5, 5, 128)  & (1, 1, 1)  \\\hline
mean  & None &($\nbatch$, 128)   & None  \\\hline
fully connected & None &($\nbatch$, 1) & None  \\\hline\hline
\end{tabular}
\caption[Architecture of the single-input CNN used in this work]{
Architecture of the single-input CNN used in this work, following \citet{Mao2021}. Tensor shapes are displayed in the order \texttt{[batch size, depth, height, width, channels]}; the PyTorch implementation internally places the channel dimension immediately after the batch dimension. Zero-padding is applied in layers \texttt{conv1}--\texttt{conv6}, and every convolutional layer is followed by a ReLU activation \citep{Nair2010}. The final feature map is spatially averaged and passed to a fully connected layer to produce one scalar prediction.}
\label{tab:cnn_single}
\end{table}

For each target position, our single-input network receives a $39\times39\times39$-voxel density sub-box centred on that position and returns one scalar: the predicted $z=10$ density at the central voxel. The resulting mapping is
\begin{align}
F:\mathbb{R}^{39\times39\times39}\rightarrow\mathbb{R}.
\end{align}
The physical size represented by the $39^3$ voxels is denoted by $\lsub$ and is varied in the analysis below.

The architecture follows \citet{Mao2021} and is summarized in Table~\ref{tab:cnn_single}. It contains seven convolutional layers. The first six use $3\times3\times3$ kernels, while the final layer uses a $1\times1\times1$ kernel. Convolutions with stride 2 reduce the spatial dimensions after the first, third, and fifth layers; no separate pooling layers are used. Zero-padding is applied in the first six convolutional layers to avoid excessive reduction of the feature maps. After the final convolution, the features are averaged over the remaining spatial dimensions and passed to a fully connected layer that produces the scalar prediction.

Our pointwise formulation differs from the map-to-map network of \citet{Shallue2023}, which implements
\begin{align}
F:\mathbb{R}^{n\times n\times n}
\rightarrow
\mathbb{R}^{(n-18)\times(n-18)\times(n-18)}.
\end{align}
Although their network predicts a region rather than a single voxel, each output voxel depends on an effective receptive field of only $17^3$ input voxels because of the finite network depth; see Section~3.1 of \citet{Shallue2023}. Thus, both architectures infer the target density primarily from information within a finite neighbourhood, even though they differ in their output format and training procedure.

In \citet{Nakashima2025MNRAS}, we varied the physical input size over $\lsub\sim38$--$380\hmpc$ while keeping the numerical input size fixed at $39^3$ voxels. This setup quantifies the trade-off between a wider physical context and a finer spatial resolution. In the present work, we repeat the same scale study after applying standard reconstruction with \texttt{nbodykit} \citep{Hand2018AJ}. This comparison tests whether removing large-scale bulk displacements changes the spatial range that the CNN must access. For reference, the $17^3$-voxel receptive field of \citet{Shallue2023}, combined with their cell size of $\lcell=3.5\hmpc$, corresponds to a physical scale of approximately $60\hmpc$, which lies within the range considered here.

We also consider the dual-input CNN introduced in \citet{Nakashima2025MNRAS}. This model receives two co-centred $39^3$ sub-boxes representing different physical scales. Each input is processed by a separate seven-layer convolutional branch, and the resulting features are combined before the final prediction; see Table~3 of \citet{Nakashima2025MNRAS}. The dual-input results without standard reconstruction are taken directly from that work. We also trained the dual-input architecture using standard-reconstructed density fields. Because it did not provide a significant improvement over the corresponding single-input hybrid model, we do not show its detailed results below.

\begin{table}
\centering
\resizebox{0.48\textwidth}{!}{
\begin{tabular}{ccccc}
\hline 
\multicolumn{2}{c|}{Input density ($z=0$)} && \multicolumn{2}{c}{Target density ($z=10$)}\\ \hline
$\lsub~[\hmpc]$ & $\lcell_\mathrm{sub}~[\hmpc]$ &$\nrebin$& $\lcell_\mathrm{box}~[\hmpc]$ & $\ngrid^3$ \\\hline \hline
38 & 0.97 &1& 0.97 & $1024^3$ \\\hline
76 & 1.95 &1& 1.95 & $512^3$ \\\hline
114 & 2.92 &3& 0.97 & $1024^3$ \\\hline
152 & 3.90 &1& 3.90 & $256^3$ \\\hline
190 & 4.88 &5& 0.97 & $1024^3$ \\\hline
228 & 5.85 &3& 1.95 & $512^3$ \\\hline
266 & 6.83 &7& 0.97 & $1024^3$ \\\hline
304 & 7.81 &1& 7.81 & $128^3$ \\\hline
342 & 8.78 &9& 0.97 & $1024^3$ \\\hline
380 & 9.76 &5& 1.95 & $512^3$ \\\hline\hline
\end{tabular}
}
\caption{Physical sizes and grid configurations used to construct the CNN inputs and targets. The parent density fields are generated in a $1\hgpc$ simulation box with grid size $\ngrid^3$ and cell size $\lcell_{\rm box}=L_{\rm box}/\ngrid$. The $z=0$ input field is rebinned by averaging over blocks of $\nrebin^3$ parent-grid cells, giving an input-cell size $\lcell_{\rm sub}=\nrebin\times\lcell_{\rm box}$. A $39^3$-voxel sub-box therefore spans the physical size $\lsub=39\times\lcell_{\rm sub}$. The target is the $z=10$ density at the central position, evaluated on the parent grid after Gaussian smoothing as described in Section~\ref{subsec:dataset}.
}
 \label{tab:cnn_scales}
\end{table}

\begin{figure*}
 \begin{tabular}{c|c}
\includegraphics[width=0.45\linewidth]{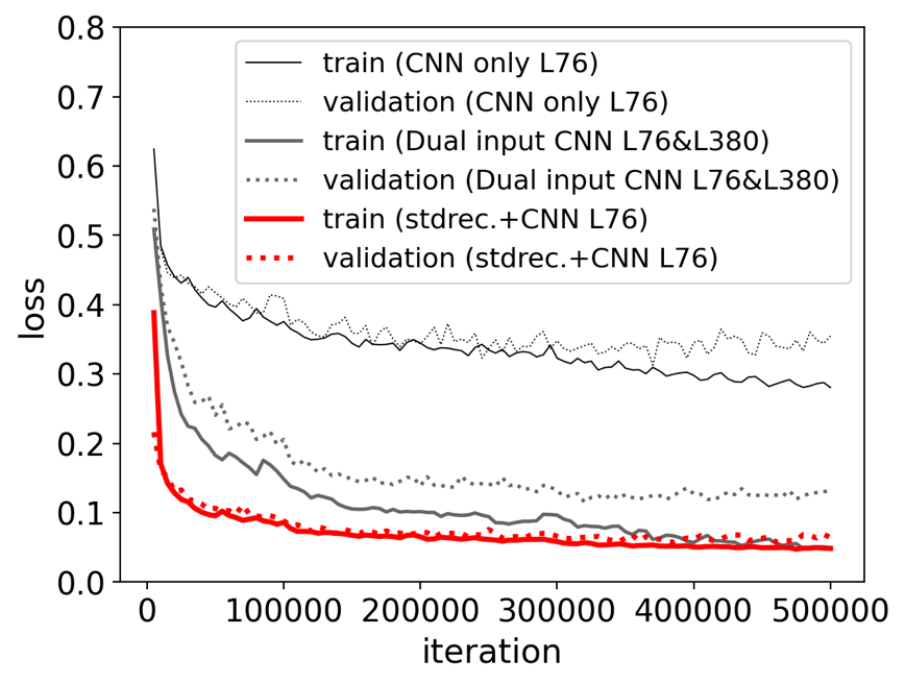} &
\includegraphics[width=0.45\linewidth]{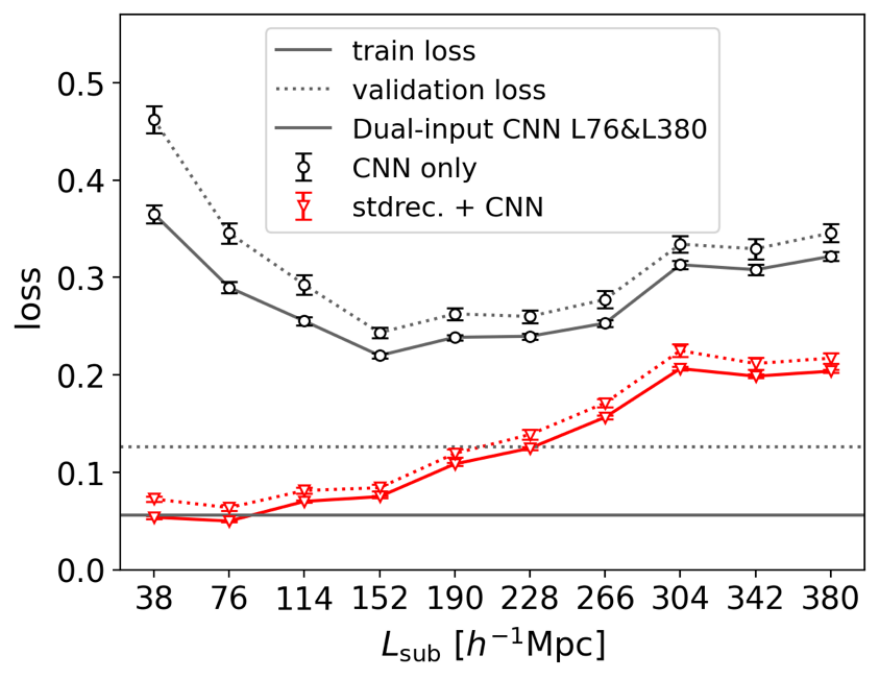}\\
 \end{tabular}
 \caption{\textit{Left}: Training and validation losses, defined by equation~(\ref{eq:loss_function}), for three representative models: the single-input CNN with $\lsub\sim76\hmpc$, corresponding to the setup of \citet{Mao2021}; the dual-input CNN with $\lsub\sim\{76,380\}\hmpc$, taken from \citet{Nakashima2025MNRAS}; and the single-input CNN with $\lsub\sim76\hmpc$ applied after standard reconstruction. Training and validation curves are distinguished by the line styles indicated in the legend. 
 \textit{Right}: Mean training and validation losses as functions of the single-input sub-box size $\lsub$. The markers show the mean loss over iterations $[4,5]\times10^5$, and the error bars show its standard deviation over the same interval. Circles denote the CNN applied directly to the $z=0$ density field (``CNN only''), while triangles denote the CNN applied after standard reconstruction (''stdrec.+CNN''). Horizontal lines show the corresponding mean losses of the dual-input CNN.}
 \label{fig:loss}
\end{figure*}

\subsection{Data set}
\label{subsec:dataset}
We use the Indra simulations~\citep{Falck2021}, a suite of large-volume $N$-body runs with WMAP7 cosmology (\(\Omega_m = 0.272\), \(\Omega_\Lambda = 0.728\), \(\Omega_b = 0.045\), \(h = 0.704\), \(\sigma_8 = 0.81\), \(n_s = 0.967\);~\citealt{Komatsu2011}) evolved with \textsc{L-Gadget2}~\citep{Springel2005}. Each realization contains $1024^3$ particles in a periodic $L_\mathrm{box}=1\hgpc$ box.
We reconstruct the dark matter density at $z=10$ from the density field at $z=0$. 

We reconstruct the dark matter density at $z=10$ from the density field at $z=0$. Except when standard reconstruction is applied to the $z=0$ particle snapshot, particle positions are assigned to regular grids using the Cloud-in-Cell (CIC) scheme.
The resulting gridded density contrast is defined as
\begin{align}
\delta
\equiv
\frac{n_{\rm DM}}{\bar{n}_{\rm DM}}-1,
\end{align}
where $n_{\rm DM}$ is the number density of dark matter particles in a grid cell and $\bar{n}_{\rm DM}$ is its spatial mean.
For the hybrid model, the $z=0$ particle snapshot is instead processed using the standard reconstruction implemented in \texttt{nbodykit} \citep{Hand2018AJ}, with a smoothing scale of $R=20\hmpc$. The output of standard reconstruction is then assigned to a regular grid and used to construct the CNN input field.

For each target position, the CNN input is a $39^3$-voxel sub-box centred on that position. In the CNN-only model, the sub-box is extracted from the original $z=0$ density field. In the hybrid model, it is extracted from the standard-reconstructed field. In both cases, the target is the $z=10$ density at the central voxel. 

To determine how much spatial context is required, we vary the physical side length of the input sub-box over $\lsub\sim38$--$380\hmpc$. Directly extracting these different volumes from a single fixed-resolution grid would produce inputs with different numbers of voxels. Instead, we keep the input dimensions fixed at $39^3$ and vary the physical cell size, as summarized in Table~\ref{tab:cnn_scales}. 
We first construct density fields in the simulation box with $L_{\rm box}=1\hgpc$ using parent grids with $\ngrid^3=128^3$--$1024^3$. The parent-grid cell size is
\begin{align}
\lcell_{\rm box}=\frac{L_{\rm box}}{\ngrid}.
\end{align}
Because the $z=10$ target density fields are constructed on parent grids with different values of $\ngrid$, their native spatial resolutions, $\lcell_{\rm box}$, also differ among the input-scale configurations. These resolution differences could affect the reconstruction accuracy and thereby complicate a direct comparison between models. To mitigate this effect, we smooth all target density fields with a Gaussian kernel of the same physical width, $3\hmpc$. This common smoothing suppresses fluctuations below approximately the same physical scale and should therefore make the reconstruction performances of the different configurations more directly comparable, although it may not completely eliminate the effects of the differing grid resolutions.

For a chosen rebinning factor $\nrebin$, the $z=0$ input field is averaged over blocks of $\nrebin^3$ neighbouring parent-grid cells. The resulting input-cell size is
\begin{align}
\lcell_{\rm sub}
=\nrebin\times\lcell_{\rm box},
\end{align}
and the physical side length of a $39^3$ input is therefore $\lsub=39\times\lcell_{\rm sub}$. This construction allows every network to receive the same number of voxels while probing different physical volumes. Increasing $\lsub$ gives the CNN access to a wider spatial environment, but at the cost of a coarser input resolution and stronger suppression of fluctuations below $\lcell_{\rm sub}$. The configuration with $\lsub\sim76\hmpc$ is identical to that used by \citet{Mao2021}. For the dual-input model, we use scale pair \(\lsub \sim\{76, 380\}\hmpc\).  Both sub-boxes are constructed from co-centred fields on a common $512^3$ parent grid so that their central positions are exactly aligned.

Training sub-boxes are sampled at intervals of $\ngrid/32$ parent-grid cells along each axis, giving $32^3=32,768$ samples per realization. We use eight realizations for training, resulting in $262,144$ distinct sub-box positions. Rotations and reflections increase the effective number of training examples by a factor of 48. The validation set contains 4096 sub-boxes extracted from a separate realization and is not augmented. Final performance is evaluated using another independent realization generated with the same cosmological parameters but different initial conditions. For this test realization, predictions are made densely at all $\ngrid^3$ grid positions without data augmentation.

\subsection{Training and Validation}
\label{subsec:train}
The network architecture and training hyperparameters follow \citet{Mao2021}; we do not perform an additional hyperparameter search. The model is implemented in PyTorch \citep{Paszke2019}. Network weights are initialized with the Xavier scheme \citep{Glorot2010} and optimized using Adam \citep{Kingma2014}.

In our training procedure, the dataset is divided into \textit{batches}, where each batch consists of a subset of the full training set. The number of samples in each batch is referred to as the batch size, denoted as $\nbatch$, leading to the total number of batches equal to 262,144$\,\times\,48/\nbatch$.
The loss function is the mean squared error (MSE), defined as
\begin{align}
\label{eq:mse}
\mathrm{MSE} = \frac{1}{\nbatch}\sum_{k=1}^{\nbatch}\left(F(\delta_\mathrm{input}^k;\bm{\theta})-\delta_\mathrm{target}^k\right)^2,
\end{align}
where $\delta_\mathrm{input}^k$ represents the input sub-box of size $39^3$ for batch element $k$, and $\delta_\mathrm{target}^k$ denotes the corresponding scalar density at $z=10$.
The MSE quantifies the average squared difference between the output values $F(\delta_\mathrm{input}^k;\bm{\theta})$ with a parameter set $\bm{\theta}$ and the target values $\delta_\mathrm{target}^k$ for each batch. In supervised learning, the loss function plays a central role in training by quantifying the discrepancy between the network's predictions and the ground truth. It serves as the primary indicator of model performance and is used to guide the optimization of trainable parameters through gradient-based updates.

We define the \textit{iteration} as a single update step of the model parameters using a batch of training data. During each iteration, the CNN processes a batch of input and target density fields, computes the loss function, and updates the model weights via back propagation using the optimizer. 
Following \citet{Mao2021}, we train each network for a fixed total of $5\times10^5$ iterations. Rather than adopting a learning-rate schedule, we keep the learning rate fixed at $10^{-4}$ throughout training and progressively increase the batch size. Specifically, we use $\nbatch=32$ for iterations $[0,1]\times10^5$, $\nbatch=128$ for $[1,3]\times10^5$, and $\nbatch=512$ for $[3,5]\times10^5$.
We monitor the MSE on both the training and validation sets. Because the validation data are not used to update the model parameters, the validation loss provides an independent measure of generalization. A decrease in the training loss accompanied by an increase in the validation loss would indicate overfitting. Nevertheless, following the training procedure of \citet{Mao2021}, we do not apply early stopping or any other validation-based termination criterion. Training is terminated after the prescribed $5\times10^5$ iterations, irrespective of the validation-loss evolution.

\begin{figure*}
\centering
    \centering
    \includegraphics[width=0.9\linewidth]{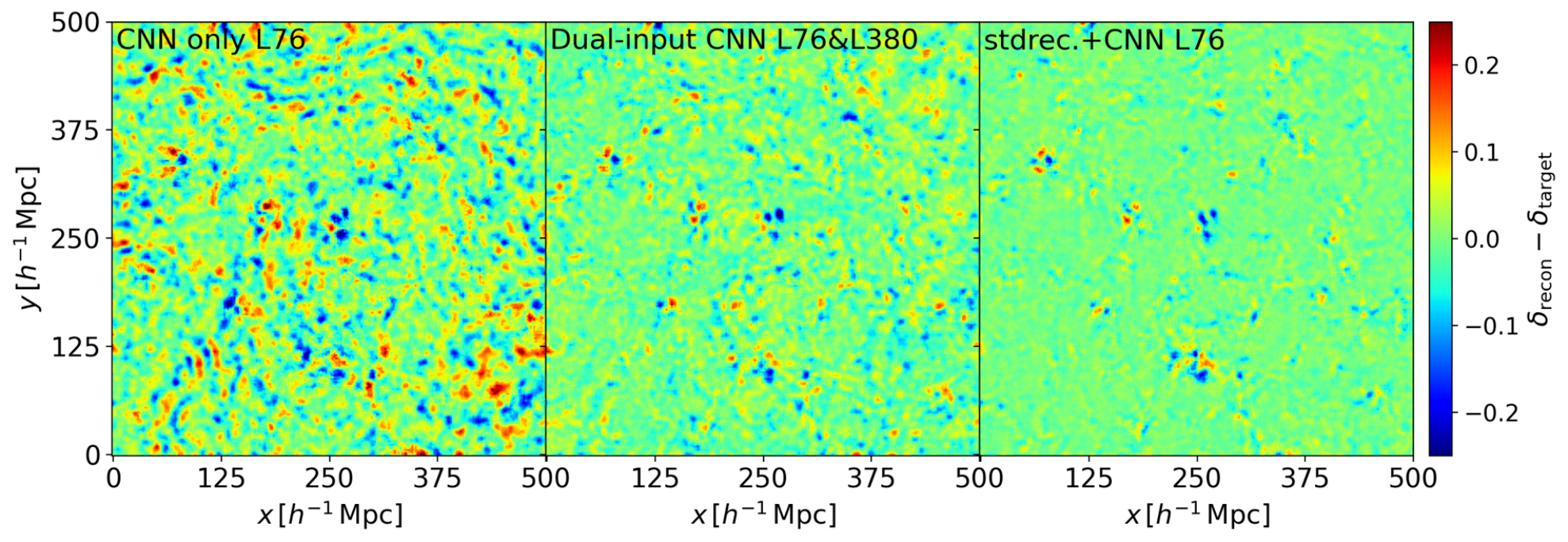}
    \caption{Slices through the residual field $\delta_{\rm recon}-\delta_{\rm target}$, with a slice thickness of $1.95\hmpc$, for three reconstruction methods. From left to right, the panels show the single-input CNN with $\lsub\sim76\hmpc$, the dual-input CNN with $\lsub\sim\{76,380\}\hmpc$, and the single-input CNN with $\lsub\sim76\hmpc$ applied after standard reconstruction. 
    }
    \label{fig:residual_maps}
\end{figure*}

\begin{figure*}
\centering
    \centering
    \includegraphics[width=0.9\linewidth]{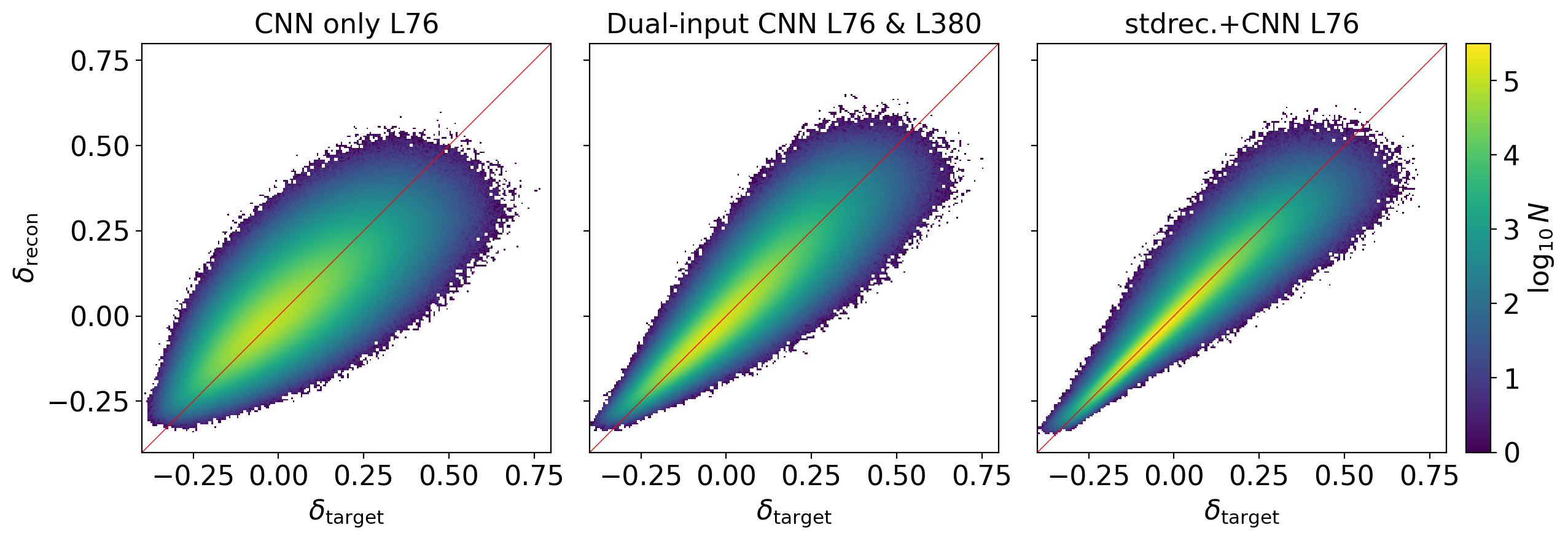}
    \caption{Two-dimensional histograms of $\dini$ versus $\drec$ for the test data set. \textit{Left}: the single-input CNN with $\lsub \sim 76\hmpc$; \textit{Centre}: the dual-input CNN with $\lsub \sim \{76,380\}\hmpc$; \textit{Right}: the single-input CNN with $\lsub \sim 76\hmpc$ applied after standard reconstruction. The colour bar represents $\log_{10}N$, where $N$ is the number of data points in each histogram bin.}

    \label{fig:contour}
\end{figure*}

\begin{figure*}
 \begin{tabular}{c|c}
\includegraphics[width=0.45\linewidth]{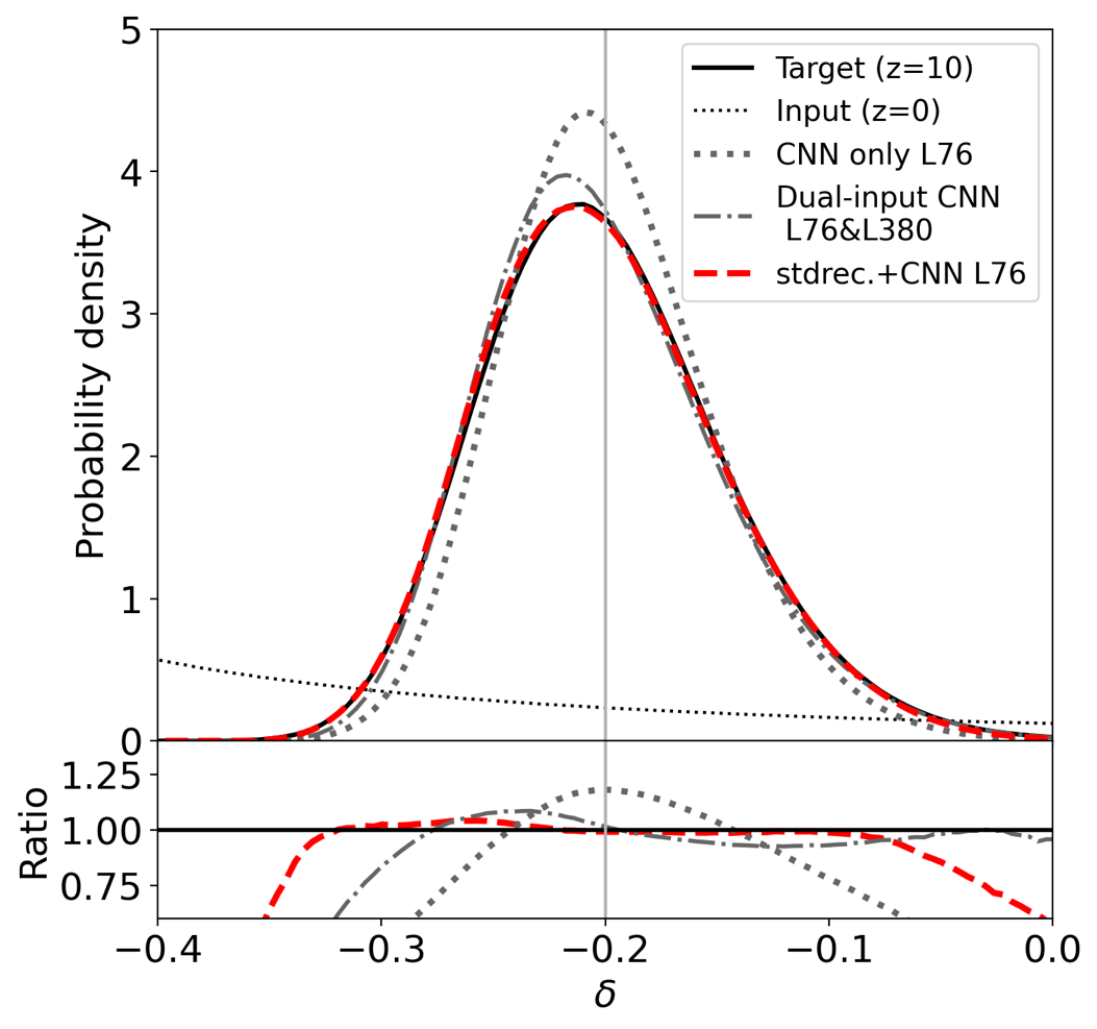} &
\includegraphics[width=0.48\linewidth]{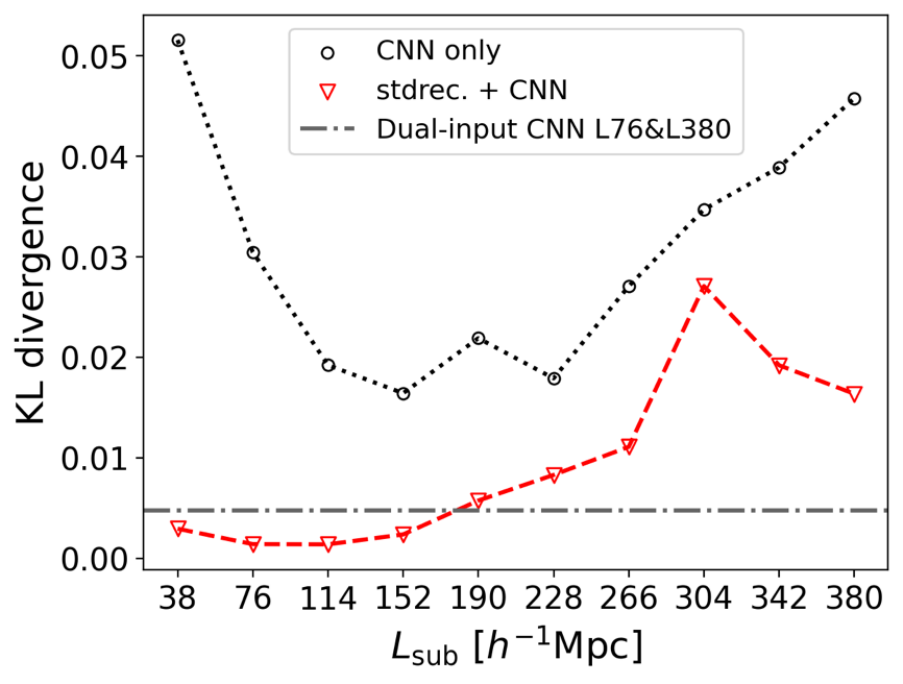}\\
 \end{tabular}
 \caption{\textit{Left}: Probability distribution functions (PDFs) of the density fluctuation $\delta$. Solid and dotted lines denote the PDFs of the target $\dini$ and input $\dfin$ density fields, respectively. The PDFs of the CNN output $\drec$ are shown for three representative setups: the single-input CNN with $\lsub\sim76\hmpc$, the dual-input CNN with $\lsub\sim\{76,380\}\hmpc$, and the single-input CNN with $\lsub\sim76\hmpc$ applied after standard reconstruction. 
 \textit{Right}: Kullback--Leibler (KL) divergence between the PDFs of the reconstructed and target density fields, defined in equation~(\ref{eq:kl_divergence}), as a function of the input sub-box size $\lsub$. Circles and triangles denote the single-input CNN applied directly to the $z=0$ density field (''CNN only'') and after standard reconstruction (''stdrec.+CNN''), respectively. The horizontal line shows the result for the dual-input CNN with $\lsub\sim\{76,380\}\hmpc$.}
 \label{fig:pdf}
\end{figure*}

\section{Results}
\label{sec:results}
We compare three reconstruction methods: a single-input CNN applied directly to the $z=0$ density field, a dual-input CNN that combines two input scales, and a single-input CNN applied to the density field after standard reconstruction. Hereafter, we refer to these methods as ``CNN only,'' ``Dual-input CNN,'' and ``stdrec.+CNN,'' respectively. The results for ``CNN only'' and ``Dual-input CNN'' are taken from \citet{Nakashima2025MNRAS}, while the results for ``stdrec.+CNN'' are obtained in this work. 

\subsection{Training loss and dependence on input scale}

In Fig.~\ref{fig:loss}, we present the loss function normalized by the variance of the target density within a batch, defined as
\begin{align}
\label{eq:loss_function}
\frac{\mathrm{MSE}}{\sigma^2} = \frac{\sum_k\left(f(\delta_\mathrm{input}^k;\bm{\theta})-\delta_\mathrm{target}^k\right)^2}{\sum_k\left(\bar{\delta}_\mathrm{target}-\delta_\mathrm{target}^k\right)^2},
\end{align}
where $k$ labels the samples within a batch and $\bar{\delta}{\rm target}$ is the mean target density contrast in that batch.

The left panel of Fig.~\ref{fig:loss} shows the evolution of the training and validation losses for three representative configurations. As reported in \citet{Nakashima2025MNRAS}, both ``CNN only'' with $\lsub\sim76\hmpc$ and the ``dual-input CNN'' with $\lsub\sim\{76,380\}\hmpc$ continue to improve over the full $5\times10^5$ training iterations. The final validation loss of the dual-input model is $0.131$, approximately $60\%$ lower than that of the single-input ``CNN only'' model. The ``stdrec.+CNN'' model with $\lsub\sim76\hmpc$ exhibits a more rapid decrease in both the training and validation losses. Its validation loss reaches $0.064$, which is approximately $80\%$ lower than that of CNN only'' and $50\%$ lower than that of the dual-input CNN. Thus, applying standard reconstruction before the CNN substantially improves the pointwise prediction of the $z=10$ density.

The right panel of Fig.~\ref{fig:loss} shows the mean loss over the final $10^5$ iterations, together with its standard deviation, as a function of the physical input size $\lsub$. The dependence on $\lsub$ differs markedly between the CNN-only and hybrid approaches. For ``CNN only,'' the validation loss decreases as the input size is increased up to $\lsub\sim152\hmpc$, beyond which the improvement saturates or reverses. 

After standard reconstruction, the trend changes. For ``stdrec.+CNN,'' the validation loss improves as the input size is reduced from large values and reaches its minimum near $\lsub\sim76\hmpc$. The dual-input result for $\lsub\sim\{76,380\}\hmpc$ is shown by the horizontal lines. Its validation loss is lower than that of ``CNN only'' for every single-input scale considered, but it is higher than that of ``stdrec.+CNN'' for $\lsub\leq190\hmpc$.
For $\lsub\sim76\hmpc$, the training loss of ``stdrec.+CNN'' is comparable to that of the dual-input CNN, while its validation loss is substantially lower. The smaller separation between the training and validation performances suggests that the hybrid model generalizes more successfully and exhibits less overfitting than the dual-input model in this configuration.

\subsection{Field-level comparison}
We perform a field-level comparison of the reconstructed density fields. Figure~\ref{fig:residual_maps} displays the residual maps ($\drec-\delta_\mathrm{target}$) for the three reconstruction methods. 
The ``CNN only'' residual field contains coherent regions of positive and negative residuals extending over relatively large spatial scales. These structures indicate that the local CNN systematically overestimates or underestimates the target density over large, spatially coherent regions. The dual-input CNN reduces the amplitudes and spatial coherence of these residuals by supplementing the small-scale input with information from a much larger sub-box.
Among the three methods, ``stdrec.+CNN'' produces the most spatially homogeneous residual field and shows the weakest coherent large-scale structures. This result is consistent with the interpretation that standard reconstruction removes a substantial part of the large-scale displacement before the density field is provided to the CNN. The CNN can therefore concentrate on the remaining, more localized reconstruction errors.

Figure~\ref{fig:contour} presents a cell-by-cell comparison between the reconstructed density $\drec$ and the target density $\dini$. All three distributions are centred approximately on the $\drec=\delta_\mathrm{target}$ relation, shown by the solid diagonal line. However, their scatter around this relation differs significantly.
The ``CNN only'' model produces the broadest distribution, consistent with the relatively large residuals visible in Fig.~\ref{fig:residual_maps}. The dual-input CNN narrows the distribution by incorporating information from both small and large physical scales. The ``stdrec.+CNN'' model produces the tightest distribution around the diagonal, indicating the smallest cell-by-cell reconstruction errors among the three methods. 

The field-level comparisons therefore support the improvement inferred from the normalized validation loss.

\begin{figure*}
 \begin{tabular}{c|c}
\includegraphics[width=0.45\linewidth]{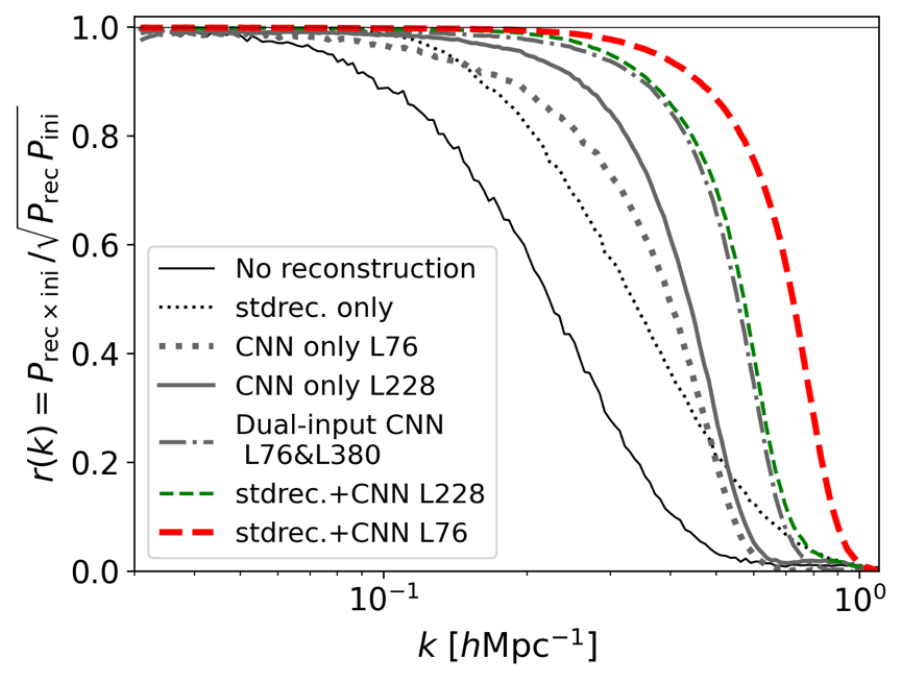} &
\includegraphics[width=0.45\linewidth]{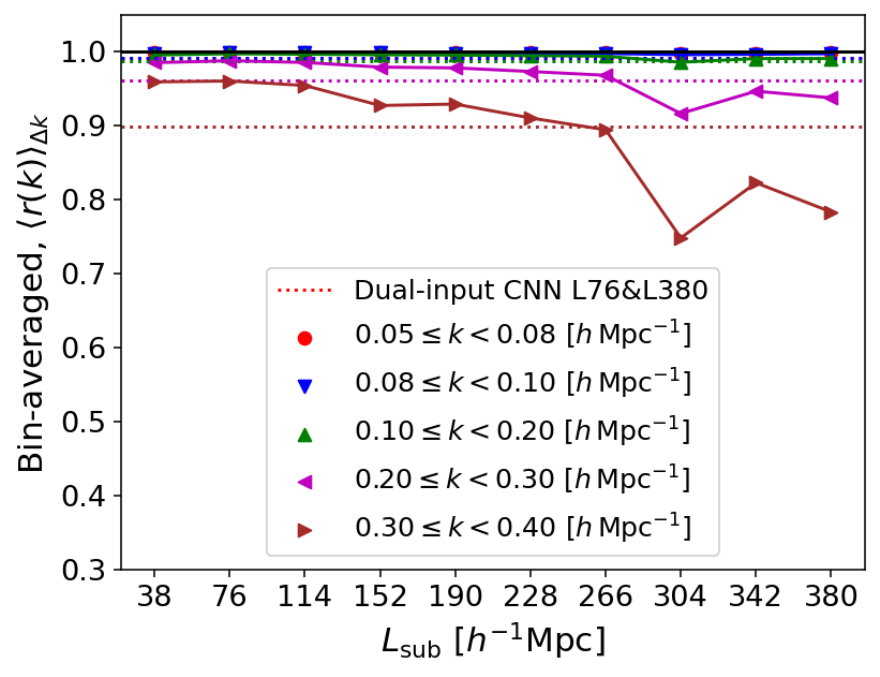}\\
 \end{tabular}
 \caption{Correlation coefficient between the reconstructed and target density fields, as defined in equation~(\ref{eq:correlation}).  \textit{Left} : Scale-dependent correlations for the different reconstruction methods. ``No reconstruction'' denotes the direct comparison between the $z=0$ and $z=10$ density fields, while ``stdrec. only'' denotes the density field obtained using standard reconstruction without a subsequent CNN.
 \textit{Right} : Correlation coefficients averaged within five wave-number intervals, shown as functions of the single-input sub-box size $\lsub$ for ``stdrec.+CNN.'' The horizontal dotted lines show the corresponding results for the dual-input CNN with $\lsub\sim\{76,380\}\hmpc$.
}
 \label{fig:corref}
\end{figure*}

\subsection{One-point density distribution}
We present the probability distribution functions (PDFs) of the density fluctuation $\delta$ and the Kullback–Leibler (KL) divergence, defined as
\begin{align}
\label{eq:kl_divergence}
{D_{\rm KL}(p \| q)=\sum_{i=1}^{N} p\left(x_{i}\right) \cdot\left[\log p\left(x_{i}\right)-\log q\left(x_{i}\right)\right],
}
\end{align}
where $p(x_i)$ and $q(x_i)$ are the normalized probabilities of the target and reconstructed density fields, respectively, in the $i$-th density bin. The KL divergence vanishes when the two distributions are identical and increases as they become more dissimilar.

The left panel of Fig.~\ref{fig:pdf} shows that ``CNN only'' with $\lsub\sim76\hmpc$ produces a narrower PDF than the target field. In particular, the model under-represents the probability of relatively large positive and negative density fluctuations. As shown in \citet{Nakashima2025MNRAS}, the dual-input CNN partially corrects this discrepancy. The ``stdrec.+CNN'' model with $\lsub\sim76\hmpc$ reproduces the target PDF more closely than either CNN-only model.

The right panel of Fig.~\ref{fig:pdf} shows the KL divergence as a function of input size $\lsub[\hmpc]$. For ``CNN only,'' the KL divergence decreases with increasing $\lsub$ and reaches its minimum at $\lsub\sim152\hmpc$. In contrast, ``stdrec.+CNN'' achieves its smallest KL divergence at $\lsub\sim76\,$--$114\hmpc$. The dual-input CNN again outperforms ``CNN only'' at every input scale considered, but its KL divergence is higher than that of ``stdrec.+CNN'' for $\lsub\leq190\hmpc$.
The shift of the minimum toward smaller $\lsub$ after standard reconstruction is consistent with the behaviour of the normalized loss. Both statistics indicate that information from a wider physical region around each target position is important when the CNN is applied directly to the evolved field. After the large-scale displacement has been partially removed, however, a smaller input volume with finer spatial resolution is preferable.

\subsection{Fourier-space correlation}
We investigate the correlation coefficient between the reconstructed and target density fields, defined as
\begin{align}
\label{eq:correlation}
    r(k)=\frac{P_\mathrm{rec\times tar}(k)}{\sqrt{P_\mathrm{rec}(k)P_\mathrm{tar}(k)}}~,
\end{align}
where $P_\mathrm{rec\times tar}(k)$ is the cross-power spectrum between the reconstructed and target density fields, and $P_\mathrm{rec}(k)$ and $P_\mathrm{tar}(k)$ are the auto-power spectra of the reconstructed and target density fields, respectively. 
Because the amplitudes are normalized by the auto-power spectra, $r(k)$ primarily measures the agreement between the spatial phases of the two fields. A perfect reconstruction gives $r(k)=1$.

The left panel of Fig.~\ref{fig:corref} compares the correlation coefficient $r(k)$ for the different reconstruction methods. In all cases, the correlation decreases toward high $k$ (small scales) due to non-linear gravitational evolution. The curve labelled ``No reconstruction'' corresponds to the direct correlation between the density fields at $z=0$ and $z=10$. It begins to deviate appreciably from unity near $k\sim0.05\ihmpc$, reflecting the accumulated non-linear displacement between the two epochs.
Standard reconstruction alone, labelled ``stdrec. only,'' maintains a higher correlation than the unreconstructed $z=0$ field over a broad range of scales. In particular, its strong performance at low $k$ demonstrates that the Zel'dovich displacement estimate successfully recovers part of the large-scale phase information.
The ``CNN only'' model with $\lsub\sim76\hmpc$ improves the correlation relative to the unreconstructed field and outperforms ``stdrec. only'' over approximately $0.15\lesssim k\lesssim0.5\ihmpc$. Standard reconstruction alone nevertheless performs better on larger scales, as previously discussed by \citet{Mao2021}. Increasing the single-input scale to $\lsub\sim228\hmpc$ or combining $\lsub\sim\{76,380\}\hmpc$ in the dual-input architecture further improves the CNN-only reconstruction, as shown in \citet{Nakashima2025MNRAS}.

For ``stdrec.+CNN,'' we show two representative input sizes, $\lsub\sim228\hmpc$ and $76\hmpc$. The model with $\lsub\sim228\hmpc$ achieves a correlation comparable to that of the dual-input CNN. Reducing the hybrid-model input to $\lsub\sim76\hmpc$ leads to a further improvement and gives the highest correlation among the methods shown over most of the resolved wave-number range.

The right panel of Fig.~\ref{fig:corref} shows the dependence on input size by averaging $r(k)$ within five wave-number intervals. In \citet{Nakashima2025MNRAS}, we found that $\lsub\sim200\hmpc$ was the optimal input size for the ''CNN only`` model. For stdrec.+CNN,'' the preferred scale is substantially smaller: configurations with $\lsub\sim38$--$76\hmpc$ generally produce the highest correlations. The precise optimal value depends on the wave-number interval, but all five intervals show the same overall shift toward smaller input sizes after standard reconstruction.

We also applied the dual-input CNN to the density field after standard reconstruction. Across the input scales and statistical measures considered here, this configuration does not produce a measurable improvement over the corresponding single-input ``stdrec.+CNN'' model. This result suggests that the additional large-scale input becomes largely redundant once standard reconstruction has already corrected the dominant large-scale displacements.

\section{Discussion}
\label{sec:discussions}
The results reveal a systematic change in the spatial range that is most informative for the CNN when standard reconstruction is applied before machine-learning reconstruction. They also clarify the complementary roles played by perturbative reconstruction and localized neural-network models.

\subsection{Shift in the preferred CNN input scale}
For ``CNN only,'' the reconstruction improves as the physical input size is increased and reaches its best performance at $\lsub\sim150$--$200\hmpc$, depending on the statistic considered. Because the numerical input dimension is fixed at $39^3$, changing $\lsub$ simultaneously changes both the physical field of view and the input-cell size. The preferred $\lsub$ should therefore not be interpreted as a pure measurement of the required physical scales. It reflects a trade-off between access to large-scale context and preservation of small-scale information. 

The preferred input size shifts to substantially smaller values after standard reconstruction. The normalized loss and KL divergence favour $\lsub\simeq76$--$114\hmpc$, whereas the Fourier-space correlation is generally highest for $\lsub\simeq38$--$76\hmpc$. Taken together, these statistics indicate a preferred range of approximately $\lsub\simeq38$--$114\hmpc$, corresponding to effective cell sizes of approximately $1$--$3\hmpc$. These finer spatial resolutions retain more of the quasi-linear cosmic-web structure in the CNN input.

This shift can be understood from the role of standard reconstruction. As discussed in Section~\ref{subsec:stdrec}, the variance of the Zel'dovich displacement is dominated by modes around $k\sim0.01\hmpc$, corresponding to wavelengths of several hundred $h^{-1}\mathrm{Mpc}$. The bulk displacement affecting the density at a given position is therefore sourced by fluctuations over a physical region much larger than the finite input volume of the CNN. When the CNN is applied directly to the evolved density field, increasing $\lsub$ allows it to access a larger fraction of this long-wavelength information. Standard reconstruction, by contrast, estimates these non-local displacements from the late-time density field and partially reverses the associated bulk flows before the field is passed to the CNN. The subsequent CNN therefore has less need for a large physical input volume and instead benefits from the finer spatial resolution provided by smaller values of $\lsub$. 

\subsection{Complementary roles of standard reconstruction and CNNs}

Our results indicate that large-scale displacement information is recovered more effectively by standard reconstruction than by the CNN architectures considered here. Standard reconstruction alone produces a higher correlation with the target field than the $\lsub\sim76\hmpc$ CNN on large scales. Moreover, ``stdrec.+CNN'' with a single $\lsub\sim76\hmpc$ input outperforms the dual-input CNN that combines $\lsub\sim76\hmpc$ and $380\hmpc$ fields.

The lack of a measurable improvement when the dual-input architecture is applied after standard reconstruction provides further support for this interpretation. Before standard reconstruction, the large input branch supplies information that is absent from the smaller local input and improves the reconstruction. After standard reconstruction, however, much of the long-wavelength displacement information has already been incorporated through the Zel'dovich estimate. The large-scale CNN branch therefore contains less additional information and becomes largely redundant.

These findings support a division of roles between the two methods. Perturbative reconstruction efficiently models coherent displacements generated by long-wavelength modes, for which the underlying gravitational dynamics are comparatively simple and physically well motivated. The CNN is then used to model the remaining relationship between the reconstructed late-time field and the target high-redshift density, including residual quasi-linear evolution and more localized non-linear structures.

\subsection{Relation to previous hybrid reconstruction studies}
Our interpretation is consistent with the hybrid method proposed by \citet{Shallue2023}. They emphasized that gravitational evolution is intrinsically non-local, whereas the prediction at each output voxel of a finite-depth CNN depends on a limited receptive field. To address this mismatch, they first applied standard reconstruction and then used a CNN to model the remaining corrections.

In their implementation, the effective receptive field spans $17^3$ voxels. With a cell size of $3.5\hmpc$, this corresponds to a physical scale of approximately $60\hmpc$. This value lies within the preferred input-scale range found for ``stdrec.+CNN'' in our analysis. Our systematic scan over $\lsub$ therefore provides an independent explanation for why a receptive scale of this order is effective after standard reconstruction.

The agreement also indicates that the performance of the hybrid method is not determined solely by combining two individually successful algorithms. Its effectiveness is connected to a physically motivated separation of scales. Standard reconstruction accounts for coherent large-scale transport, whereas the CNN operates on a field in which the remaining reconstruction problem is more local.

\section{Conclusions}
\label{sec:conclusions}
In this work, we investigated how the physical scale of the CNN input and the choice of reconstruction strategy affect the recovery of the dark matter density field at $z=10$. We compared three approaches: a single-input CNN applied directly to the evolved density field (``CNN only''), a dual-input CNN combining two physical scales (``dual-input CNN''), and a single-input CNN applied after standard reconstruction (``stdrec.+CNN''). Their performance was evaluated using the normalized loss, residual fields, one-point density distributions, KL divergence, and Fourier-space correlation coefficients.

For ``CNN only,'' the reconstruction performs best for input sizes of approximately $\lsub\sim150$--$200\hmpc$. This preference for relatively large input volumes indicates that the CNN benefits from access to long-wavelength information associated with coherent matter displacements. In contrast, after standard reconstruction, the preferred input scale shifts to $\lsub\sim38$--$114\hmpc$, as consistently indicated by the loss, KL divergence, and correlation statistics. Standard reconstruction estimates and partially reverses large-scale bulk flows before the density field is passed to the CNN, thereby reducing the need for the network to infer these non-local displacements from a finite input region. The subsequent CNN can instead operate on a smaller input volume with finer spatial resolution and focus on the remaining quasi-linear and non-linear evolution.

The ``stdrec.+CNN'' approach achieves better reconstruction performance than both the single- and dual-input CNNs across the statistical measures considered here. We also find no measurable improvement when the dual-input architecture is applied after standard reconstruction, suggesting that its additional large-scale branch becomes largely redundant once the dominant large-scale displacements have already been corrected. These results support a complementary division of roles: perturbative methods based on the Zel'dovich approximation efficiently recover coherent long-wavelength displacements, whereas CNNs are more effective at modelling the residual evolution on intermediate and smaller scales.

The preferred scale found for ``stdrec.+CNN'' is also consistent with the effective receptive scale of approximately $60\hmpc$ in the hybrid model of \citet{Shallue2023}. This agreement provides further support for a physically motivated separation of scales in cosmological reconstruction. Rather than relying solely on increasingly large receptive fields or more complex neural-network architectures, a hybrid framework can combine analytic reconstruction and machine learning so that each method addresses the scales for which it is best suited.

Several extensions will be important for assessing the applicability of these results. First, the preferred input scales identified here apply to idealized real-space dark matter fields. Redshift-space distortions introduce anisotropic displacements and alter the distribution of information across spatial scales, and may therefore change both the optimal input size and the relative roles of standard reconstruction and the CNN. Second, observational effects such as tracer bias, shot noise, survey geometry, and incomplete volume coverage must be incorporated before the method can be applied to galaxy surveys. Third, alternative architectures, including U-Nets or models with explicitly non-local operations, may improve the reconstruction of residual structures that remain difficult for a local CNN. Finally, higher-order perturbative schemes and iterative reconstruction methods may provide a more accurate treatment of quasi-linear evolution and help clarify how the division of roles between analytic and data-driven reconstruction changes as the perturbative component is improved.


\section*{Acknowledgements}
We thank Luisa Lucie-Smith for helpful comments and valuable suggestions, and Kenji Hasegawa for insightful discussions that contributed to the development of this work.
This work is supported by JSPS Kakenhi Grant Numbers: JP22K21349, JP23H00108, 25H01551 (AJN), 21H04467, and 24K00625 (KI). It is also supported by the JST FOREST Program JPMJFR20352935 and the JSPS Core-to-Core Program (grant numbers: JPJSCCA20200002 and JPJSCCA20200003).

\section*{Data Availability}
The data underlying this article will be shared on reasonable request to the corresponding author.



\bibliographystyle{mnras}
\bibliography{reference} 

@ARTICLE{2018MNRAS.478.1866H,
       author = {{Hada}, Ryuichiro and {Eisenstein}, Daniel J.},
        title = "{An iterative reconstruction of cosmological initial density fields}",
      journal = {\mnras},
         year = 2018,
        month = aug,
       volume = {478},
       number = {2},
        pages = {1866-1874},
          doi = {10.1093/mnras/sty1203},
archivePrefix = {arXiv},
       eprint = {1804.04738},
 primaryClass = {astro-ph.CO},
       adsurl = {https://ui.adsabs.harvard.edu/abs/2018MNRAS.478.1866H}
}

@ARTICLE{2022PhRvD.105f3507I,
       author = {{Ichiki}, Kiyotomo and {Sumiya}, Kento and {Liu}, Guo-Chin},
        title = "{Measuring the cosmological density field twice: A novel test of dark energy using the CMB quadrupole}",
      journal = {\prd},
         year = 2022,
        month = mar,
       volume = {105},
       number = {6},
          eid = {063507},
        pages = {063507},
          doi = {10.1103/PhysRevD.105.063507},
archivePrefix = {arXiv},
       eprint = {2202.11332},
 primaryClass = {astro-ph.CO},
       adsurl = {https://ui.adsabs.harvard.edu/abs/2022PhRvD.105f3507I}
}

@ARTICLE{Mao2021,
       author = {{Mao}, Tian-Xiang and {Wang}, Jie and {Li}, Baojiu and {Cai}, Yan-Chuan and {Falck}, Bridget and {Neyrinck}, Mark and {Szalay}, Alex},
        title = "{Baryon acoustic oscillations reconstruction using convolutional neural networks}",
      journal = {\mnras},
         year = 2021,
        month = feb,
       volume = {501},
       number = {1},
        pages = {1499-1510},
          doi = {10.1093/mnras/staa3741},
archivePrefix = {arXiv},
       eprint = {2002.10218},
 primaryClass = {astro-ph.CO},
       adsurl = {https://ui.adsabs.harvard.edu/abs/2021MNRAS.501.1499M}
}

@ARTICLE{Falck2021,
       author = {{Falck}, Bridget and {Wang}, Jie and {Jenkins}, Adrian and {Lemson}, Gerard and {Medvedev}, Dmitry and {Neyrinck}, Mark C. and {Szalay}, Alex S.},
        title = "{Indra: a public computationally accessible suite of cosmological N-body simulations}",
      journal = {\mnras},
         year = 2021,
        month = sep,
       volume = {506},
       number = {2},
        pages = {2659-2670},
          doi = {10.1093/mnras/stab1823},
archivePrefix = {arXiv},
       eprint = {2101.03631},
 primaryClass = {astro-ph.CO},
       adsurl = {https://ui.adsabs.harvard.edu/abs/2021MNRAS.506.2659F}
}

@ARTICLE{Springel2005,
       author = {{Springel}, Volker},
        title = "{The cosmological simulation code GADGET-2}",
      journal = {\mnras},
         year = 2005,
        month = dec,
       volume = {364},
       number = {4},
        pages = {1105-1134},
          doi = {10.1111/j.1365-2966.2005.09655.x},
archivePrefix = {arXiv},
       eprint = {astro-ph/0505010},
 primaryClass = {astro-ph},
       adsurl = {https://ui.adsabs.harvard.edu/abs/2005MNRAS.364.1105S}
}

@ARTICLE{Komatsu2011,
       author = {{Komatsu}, E. and {Smith}, K.~M. and {Dunkley}, J. and {Bennett}, C.~L. and {Gold}, B. and {Hinshaw}, G. and {Jarosik}, N. and {Larson}, D. and {Nolta}, M.~R. and {Page}, L. and {Spergel}, D.~N. and {Halpern}, M. and {Hill}, R.~S. and {Kogut}, A. and {Limon}, M. and {Meyer}, S.~S. and {Odegard}, N. and {Tucker}, G.~S. and {Weiland}, J.~L. and {Wollack}, E. and {Wright}, E.~L.},
        title = "{Seven-year Wilkinson Microwave Anisotropy Probe (WMAP) Observations: Cosmological Interpretation}",
      journal = {\apjs},
         year = 2011,
        month = feb,
       volume = {192},
       number = {2},
          eid = {18},
        pages = {18},
          doi = {10.1088/0067-0049/192/2/18},
archivePrefix = {arXiv},
       eprint = {1001.4538},
 primaryClass = {astro-ph.CO},
       adsurl = {https://ui.adsabs.harvard.edu/abs/2011ApJS..192...18K}
}

@inproceedings{Nair2010,
author = {Nair, Vinod and Hinton, Geoffrey E.},
title = {Rectified linear units improve restricted boltzmann machines},
year = {2010},
isbn = {9781605589077},
publisher = {Omnipress},
address = {Madison, WI, USA},
booktitle = {Proceedings of the 27th International Conference on International Conference on Machine Learning},
pages = {807–814},
numpages = {8},
location = {Haifa, Israel},
series = {ICML'10}
}

@inproceedings{Krizhevsky2012,
author = {Krizhevsky, Alex and Sutskever, Ilya and Hinton, Geoffrey E.},
title = {ImageNet classification with deep convolutional neural networks},
year = {2012},
publisher = {Curran Associates Inc.},
address = {Red Hook, NY, USA},
booktitle = {Proceedings of the 26th International Conference on Neural Information Processing Systems - Volume 1},
pages = {1097–1105},
numpages = {9},
location = {Lake Tahoe, Nevada},
series = {NIPS'12}
}

@ARTICLE{Kingma2014,
       author = {{Kingma}, Diederik P. and {Ba}, Jimmy},
        title = "{Adam: A Method for Stochastic Optimization}",
      journal = {arXiv e-prints},
         year = 2014,
        month = dec,
          eid = {arXiv:1412.6980},
        pages = {arXiv:1412.6980},
          doi = {10.48550/arXiv.1412.6980},
archivePrefix = {arXiv},
       eprint = {1412.6980},
 primaryClass = {cs.LG},
       adsurl = {https://ui.adsabs.harvard.edu/abs/2014arXiv1412.6980K}
}

@InProceedings{Glorot2010,
  title = 	 {Understanding the difficulty of training deep feedforward neural networks},
  author = 	 {Glorot, Xavier and Bengio, Yoshua},
  booktitle = 	 {Proceedings of the Thirteenth International Conference on Artificial Intelligence and Statistics},
  pages = 	 {249--256},
  year = 	 {2010},
  editor = 	 {Teh, Yee Whye and Titterington, Mike},
  volume = 	 {9},
  series = 	 {Proceedings of Machine Learning Research},
  address = 	 {Chia Laguna Resort, Sardinia, Italy},
  month = 	 {13--15 May},
  publisher =    {PMLR},
  url = 	 {https://proceedings.mlr.press/v9/glorot10a.html}
}

@ARTICLE{Paszke2019,
       author = {{Paszke}, Adam and {Gross}, Sam and {Massa}, Francisco and {Lerer}, Adam and {Bradbury}, James and {Chanan}, Gregory and {Killeen}, Trevor and {Lin}, Zeming and {Gimelshein}, Natalia and {Antiga}, Luca and {Desmaison}, Alban and {K{\"o}pf}, Andreas and {Yang}, Edward and {DeVito}, Zach and {Raison}, Martin and {Tejani}, Alykhan and {Chilamkurthy}, Sasank and {Steiner}, Benoit and {Fang}, Lu and {Bai}, Junjie and {Chintala}, Soumith},
        title = "{PyTorch: An Imperative Style, High-Performance Deep Learning Library}",
      journal = {arXiv e-prints},
         year = 2019,
        month = dec,
          eid = {arXiv:1912.01703},
        pages = {arXiv:1912.01703},
          doi = {10.48550/arXiv.1912.01703},
archivePrefix = {arXiv},
       eprint = {1912.01703},
 primaryClass = {cs.LG},
       adsurl = {https://ui.adsabs.harvard.edu/abs/2019arXiv191201703P}
}

@ARTICLE{Shallue2023,
       author = {{Shallue}, Christopher J. and {Eisenstein}, Daniel J.},
        title = "{Reconstructing cosmological initial conditions from late-time structure with convolutional neural networks}",
      journal = {\mnras},
         year = 2023,
        month = apr,
       volume = {520},
       number = {4},
        pages = {6256-6267},
          doi = {10.1093/mnras/stad528},
archivePrefix = {arXiv},
       eprint = {2207.12511},
 primaryClass = {astro-ph.CO},
       adsurl = {https://ui.adsabs.harvard.edu/abs/2023MNRAS.520.6256S}
}

@ARTICLE{Kamionkowski1997,
       author = {{Kamionkowski}, Marc and {Loeb}, Abraham},
        title = "{Getting around cosmic variance}",
      journal = {\prd},
         year = 1997,
        month = oct,
       volume = {56},
       number = {8},
        pages = {4511-4513},
          doi = {10.1103/PhysRevD.56.4511},
archivePrefix = {arXiv},
       eprint = {astro-ph/9703118},
 primaryClass = {astro-ph},
       adsurl = {https://ui.adsabs.harvard.edu/abs/1997PhRvD..56.4511K}
}

@ARTICLE{Seto2000,
       author = {{Seto}, Naoki and {Sasaki}, Misao},
        title = "{Polarization signal of distant clusters and reconstruction of primordial potential fluctuations}",
      journal = {\prd},
         year = 2000,
        month = dec,
       volume = {62},
       number = {12},
          eid = {123004},
        pages = {123004},
          doi = {10.1103/PhysRevD.62.123004},
archivePrefix = {arXiv},
       eprint = {astro-ph/0009222},
 primaryClass = {astro-ph},
       adsurl = {https://ui.adsabs.harvard.edu/abs/2000PhRvD..62l3004S}
}

@ARTICLE{Weinberg2013,
       author = {{Weinberg}, David H. and {Mortonson}, Michael J. and {Eisenstein}, Daniel J. and {Hirata}, Christopher and {Riess}, Adam G. and {Rozo}, Eduardo},
        title = "{Observational probes of cosmic acceleration}",
      journal = {\physrep},
         year = 2013,
        month = sep,
       volume = {530},
       number = {2},
        pages = {87-255},
          doi = {10.1016/j.physrep.2013.05.001},
archivePrefix = {arXiv},
       eprint = {1201.2434},
 primaryClass = {astro-ph.CO},
       adsurl = {https://ui.adsabs.harvard.edu/abs/2013PhR...530...87W}
}

@ARTICLE{Portsmouth2004,
       author = {{Portsmouth}, Jamie},
        title = "{Analysis of the Kamionkowski-Loeb method of reducing cosmic variance with CMB polarization}",
      journal = {\prd},
         year = 2004,
        month = sep,
       volume = {70},
       number = {6},
          eid = {063504},
        pages = {063504},
          doi = {10.1103/PhysRevD.70.063504},
archivePrefix = {arXiv},
       eprint = {astro-ph/0402173},
 primaryClass = {astro-ph},
       adsurl = {https://ui.adsabs.harvard.edu/abs/2004PhRvD..70f3504P}
}

@ARTICLE{Bartolo2004,
       author = {{Bartolo}, N. and {Komatsu}, E. and {Matarrese}, S. and {Riotto}, A.},
        title = "{Non-Gaussianity from inflation: theory and observations}",
      journal = {\physrep},
         year = 2004,
        month = nov,
       volume = {402},
       number = {3-4},
        pages = {103-266},
          doi = {10.1016/j.physrep.2004.08.022},
archivePrefix = {arXiv},
       eprint = {astro-ph/0406398},
 primaryClass = {astro-ph},
       adsurl = {https://ui.adsabs.harvard.edu/abs/2004PhR...402..103B}
}

@ARTICLE{Nusser1992,
       author = {{Nusser}, Adi and {Dekel}, Avishai},
        title = "{Tracing Large-Scale Fluctuations Back in Time}",
      journal = {\apj},
         year = 1992,
        month = jun,
       volume = {391},
        pages = {443},
          doi = {10.1086/171360},
       adsurl = {https://ui.adsabs.harvard.edu/abs/1992ApJ...391..443N}
}

@ARTICLE{Eisenstein2007,
       author = {{Eisenstein}, Daniel J. and {Seo}, Hee-Jong and {Sirko}, Edwin and {Spergel}, David N.},
        title = "{Improving Cosmological Distance Measurements by Reconstruction of the Baryon Acoustic Peak}",
      journal = {\apj},
         year = 2007,
        month = aug,
       volume = {664},
       number = {2},
        pages = {675-679},
          doi = {10.1086/518712},
archivePrefix = {arXiv},
       eprint = {astro-ph/0604362},
 primaryClass = {astro-ph},
       adsurl = {https://ui.adsabs.harvard.edu/abs/2007ApJ...664..675E}
}

@ARTICLE{Seo2008,
       author = {{Seo}, Hee-Jong and {Siegel}, Ethan R. and {Eisenstein}, Daniel J. and {White}, Martin},
        title = "{Nonlinear Structure Formation and the Acoustic Scale}",
      journal = {\apj},
         year = 2008,
        month = oct,
       volume = {686},
       number = {1},
        pages = {13-24},
          doi = {10.1086/589921},
archivePrefix = {arXiv},
       eprint = {0805.0117},
 primaryClass = {astro-ph},
       adsurl = {https://ui.adsabs.harvard.edu/abs/2008ApJ...686...13S}
}

@ARTICLE{Noh2009,
       author = {{Noh}, Yookyung and {White}, Martin and {Padmanabhan}, Nikhil},
        title = "{Reconstructing baryon oscillations}",
      journal = {\prd},
         year = 2009,
        month = dec,
       volume = {80},
       number = {12},
          eid = {123501},
        pages = {123501},
          doi = {10.1103/PhysRevD.80.123501},
archivePrefix = {arXiv},
       eprint = {0909.1802},
 primaryClass = {astro-ph.CO},
       adsurl = {https://ui.adsabs.harvard.edu/abs/2009PhRvD..80l3501N}
}

@ARTICLE{Padmanabhan2009,
       author = {{Padmanabhan}, Nikhil and {White}, Martin and {Cohn}, J.~D.},
        title = "{Reconstructing baryon oscillations: A Lagrangian theory perspective}",
      journal = {\prd},
         year = 2009,
        month = mar,
       volume = {79},
       number = {6},
          eid = {063523},
        pages = {063523},
          doi = {10.1103/PhysRevD.79.063523},
archivePrefix = {arXiv},
       eprint = {0812.2905},
 primaryClass = {astro-ph},
       adsurl = {https://ui.adsabs.harvard.edu/abs/2009PhRvD..79f3523P}
}

@ARTICLE{Schmittfull2015,
       author = {{Schmittfull}, Marcel and {Feng}, Yu and {Beutler}, Florian and {Sherwin}, Blake and {Chu}, Man Yat},
        title = "{Eulerian BAO reconstructions and N -point statistics}",
      journal = {\prd},
         year = 2015,
        month = dec,
       volume = {92},
       number = {12},
          eid = {123522},
        pages = {123522},
          doi = {10.1103/PhysRevD.92.123522},
archivePrefix = {arXiv},
       eprint = {1508.06972},
 primaryClass = {astro-ph.CO},
       adsurl = {https://ui.adsabs.harvard.edu/abs/2015PhRvD..92l3522S}
}

@ARTICLE{Xu2013,
       author = {{Xu}, Xiaoying and {Cuesta}, Antonio J. and {Padmanabhan}, Nikhil and {Eisenstein}, Daniel J. and {McBride}, Cameron K.},
        title = "{Measuring D$_{A}$ and H at z=0.35 from the SDSS DR7 LRGs using baryon acoustic oscillations}",
      journal = {\mnras},
         year = 2013,
        month = may,
       volume = {431},
       number = {3},
        pages = {2834-2860},
          doi = {10.1093/mnras/stt379},
archivePrefix = {arXiv},
       eprint = {1206.6732},
 primaryClass = {astro-ph.CO},
       adsurl = {https://ui.adsabs.harvard.edu/abs/2013MNRAS.431.2834X}
}

@ARTICLE{Hinton2017,
       author = {{Hinton}, Samuel R. and {Kazin}, Eyal and {Davis}, Tamara M. and {Blake}, Chris and {Brough}, Sarah and {Colless}, Matthew and {Couch}, Warrick J. and {Drinkwater}, Michael J. and {Glazebrook}, Karl and {Jurek}, Russell J. and {Parkinson}, David and {Pimbblet}, Kevin A. and {Poole}, Gregory B. and {Pracy}, Michael and {Woods}, David},
        title = "{Measuring the 2D baryon acoustic oscillation signal of galaxies in WiggleZ: cosmological constraints}",
      journal = {\mnras},
         year = 2017,
        month = feb,
       volume = {464},
       number = {4},
        pages = {4807-4822},
          doi = {10.1093/mnras/stw2725},
archivePrefix = {arXiv},
       eprint = {1611.08040},
 primaryClass = {astro-ph.CO},
       adsurl = {https://ui.adsabs.harvard.edu/abs/2017MNRAS.464.4807H}
}

@ARTICLE{LeCun2015,
       author = {{LeCun}, Yann and {Bengio}, Yoshua and {Hinton}, Geoffrey},
        title = "{Deep learning}",
      journal = {\nat},
         year = 2015,
        month = may,
       volume = {521},
       number = {7553},
        pages = {436-444},
          doi = {10.1038/nature14539},
       adsurl = {https://ui.adsabs.harvard.edu/abs/2015Natur.521..436L}
}

@book{Goodfellow2016,
    title={Deep Learning},
    author={Ian Goodfellow and Yoshua Bengio and Aaron Courville},
    publisher={MIT Press},
    note={\url{http://www.deeplearningbook.org}},
    year={2016}
}

@ARTICLE{Parker2025,
       author = {{Parker}, Liam and {Bayer}, Adrian E. and {Seljak}, Uros},
        title = "{Initial Conditions from Galaxies: Machine-Learning Subgrid Correction to Standard Reconstruction}",
      journal = {arXiv e-prints},
         year = 2025,
        month = apr,
          eid = {arXiv:2504.01092},
        pages = {arXiv:2504.01092},
          doi = {10.48550/arXiv.2504.01092},
archivePrefix = {arXiv},
       eprint = {2504.01092},
 primaryClass = {astro-ph.CO},
       adsurl = {https://ui.adsabs.harvard.edu/abs/2025arXiv250401092P}
}

@ARTICLE{Chen2023,
       author = {{Chen}, Xinyi and {Zhu}, Fangzhou and {Gaines}, Sasha and {Padmanabhan}, Nikhil},
        title = "{Effective cosmic density field reconstruction with convolutional neural network}",
      journal = {\mnras},
         year = 2023,
        month = aug,
       volume = {523},
       number = {4},
        pages = {6272-6281},
          doi = {10.1093/mnras/stad1868},
archivePrefix = {arXiv},
       eprint = {2306.10538},
 primaryClass = {astro-ph.CO},
       adsurl = {https://ui.adsabs.harvard.edu/abs/2023MNRAS.523.6272C}
}

@ARTICLE{Zhu2016,
       author = {{Zhu}, Hong-Ming and {Pen}, Ue-Li and {Chen}, Xuelei},
        title = "{Primordial density and BAO reconstruction}",
      journal = {arXiv e-prints},
         year = 2016,
        month = sep,
          eid = {arXiv:1609.07041},
        pages = {arXiv:1609.07041},
          doi = {10.48550/arXiv.1609.07041},
archivePrefix = {arXiv},
       eprint = {1609.07041},
 primaryClass = {astro-ph.CO},
       adsurl = {https://ui.adsabs.harvard.edu/abs/2016arXiv160907041Z}
}

@ARTICLE{Zhu2017,
       author = {{Zhu}, Hong-Ming and {Yu}, Yu and {Pen}, Ue-Li and {Chen}, Xuelei and {Yu}, Hao-Ran},
        title = "{Nonlinear reconstruction}",
      journal = {\prd},
         year = 2017,
        month = dec,
       volume = {96},
       number = {12},
          eid = {123502},
        pages = {123502},
          doi = {10.1103/PhysRevD.96.123502},
archivePrefix = {arXiv},
       eprint = {1611.09638},
 primaryClass = {astro-ph.CO},
       adsurl = {https://ui.adsabs.harvard.edu/abs/2017PhRvD..96l3502Z}
}

@ARTICLE{Pan2017,
       author = {{Pan}, Qiaoyin and {Pen}, Ue-Li and {Inman}, Derek and {Yu}, Hao-Ran},
        title = "{Increasing Fisher information by Potential Isobaric Reconstruction}",
      journal = {\mnras},
         year = 2017,
        month = aug,
       volume = {469},
       number = {2},
        pages = {1968-1973},
          doi = {10.1093/mnras/stx774},
archivePrefix = {arXiv},
       eprint = {1611.10013},
 primaryClass = {astro-ph.CO},
       adsurl = {https://ui.adsabs.harvard.edu/abs/2017MNRAS.469.1968P}
}

@ARTICLE{Wang2017,
       author = {{Wang}, Xin and {Yu}, Hao-Ran and {Zhu}, Hong-Ming and {Yu}, Yu and {Pan}, Qiaoyin and {Pen}, Ue-Li},
        title = "{Isobaric Reconstruction of the Baryonic Acoustic Oscillation}",
      journal = {\apjl},
         year = 2017,
        month = jun,
       volume = {841},
       number = {2},
          eid = {L29},
        pages = {L29},
          doi = {10.3847/2041-8213/aa738c},
archivePrefix = {arXiv},
       eprint = {1703.09742},
 primaryClass = {astro-ph.CO},
       adsurl = {https://ui.adsabs.harvard.edu/abs/2017ApJ...841L..29W}
}

@ARTICLE{Yu2017,
       author = {{Yu}, Yu and {Zhu}, Hong-Ming and {Pen}, Ue-Li},
        title = "{Halo Nonlinear Reconstruction}",
      journal = {\apj},
         year = 2017,
        month = oct,
       volume = {847},
       number = {2},
          eid = {110},
        pages = {110},
          doi = {10.3847/1538-4357/aa89e7},
archivePrefix = {arXiv},
       eprint = {1703.08301},
 primaryClass = {astro-ph.CO},
       adsurl = {https://ui.adsabs.harvard.edu/abs/2017ApJ...847..110Y}
}

@ARTICLE{WangPen2019,
       author = {{Wang}, Xin and {Pen}, Ue-Li},
        title = "{Understanding the Reconstruction of the Biased Tracer}",
      journal = {\apj},
         year = 2019,
        month = jan,
       volume = {870},
       number = {2},
          eid = {116},
        pages = {116},
          doi = {10.3847/1538-4357/aaf231},
archivePrefix = {arXiv},
       eprint = {1807.06381},
 primaryClass = {astro-ph.CO},
       adsurl = {https://ui.adsabs.harvard.edu/abs/2019ApJ...870..116W}
}

@ARTICLE{Zhu2018,
       author = {{Zhu}, Hong-Ming and {Yu}, Yu and {Pen}, Ue-Li},
        title = "{Nonlinear reconstruction of redshift space distortions}",
      journal = {\prd},
         year = 2018,
        month = feb,
       volume = {97},
       number = {4},
          eid = {043502},
        pages = {043502},
          doi = {10.1103/PhysRevD.97.043502},
archivePrefix = {arXiv},
       eprint = {1711.03218},
 primaryClass = {astro-ph.CO},
       adsurl = {https://ui.adsabs.harvard.edu/abs/2018PhRvD..97d3502Z}
}

@ARTICLE{Schmittfull2017,
       author = {{Schmittfull}, Marcel and {Baldauf}, Tobias and {Zaldarriaga}, Matias},
        title = "{Iterative initial condition reconstruction}",
      journal = {\prd},
         year = 2017,
        month = jul,
       volume = {96},
       number = {2},
          eid = {023505},
        pages = {023505},
          doi = {10.1103/PhysRevD.96.023505},
archivePrefix = {arXiv},
       eprint = {1704.06634},
 primaryClass = {astro-ph.CO},
       adsurl = {https://ui.adsabs.harvard.edu/abs/2017PhRvD..96b3505S}
}

@ARTICLE{Seljak2017,
       author = {{Seljak}, Uro{\v{s}} and {Aslanyan}, Grigor and {Feng}, Yu and {Modi}, Chirag},
        title = "{Towards optimal extraction of cosmological information from nonlinear data}",
      journal = {\jcap},
         year = 2017,
        month = dec,
       volume = {2017},
       number = {12},
          eid = {009},
        pages = {009},
          doi = {10.1088/1475-7516/2017/12/009},
archivePrefix = {arXiv},
       eprint = {1706.06645},
 primaryClass = {astro-ph.CO},
       adsurl = {https://ui.adsabs.harvard.edu/abs/2017JCAP...12..009S}
}

@ARTICLE{Shi2018,
       author = {{Shi}, Yanlong and {Cautun}, Marius and {Li}, Baojiu},
        title = "{New method for initial density reconstruction}",
      journal = {\prd},
         year = 2018,
        month = jan,
       volume = {97},
       number = {2},
          eid = {023505},
        pages = {023505},
          doi = {10.1103/PhysRevD.97.023505},
archivePrefix = {arXiv},
       eprint = {1709.06350},
 primaryClass = {astro-ph.CO},
       adsurl = {https://ui.adsabs.harvard.edu/abs/2018PhRvD..97b3505S}
}

@ARTICLE{Wang2020,
       author = {{Wang}, Yuchan and {Li}, Baojiu and {Cautun}, Marius},
        title = "{Iterative removal of redshift-space distortions from galaxy clustering}",
      journal = {\mnras},
         year = 2020,
        month = sep,
       volume = {497},
       number = {3},
        pages = {3451-3471},
          doi = {10.1093/mnras/staa2136},
archivePrefix = {arXiv},
       eprint = {1912.03392},
 primaryClass = {astro-ph.CO},
       adsurl = {https://ui.adsabs.harvard.edu/abs/2020MNRAS.497.3451W}
}

@ARTICLE{Nakashima2025MNRAS,
       author = {{Nakashima}, Koichiro and {Ichiki}, Kiyotomo and {Nishizawa}, Atsushi J. and {Hasegawa}, Kenji},
        title = "{Searching optimal scales for reconstructing cosmological initial conditions using convolutional neural networks}",
      journal = {\mnras},
         year = 2025,
        month = dec,
       volume = {544},
       number = {2},
        pages = {2586-2598},
          doi = {10.1093/mnras/staf1802},
archivePrefix = {arXiv},
       eprint = {2505.10636},
 primaryClass = {astro-ph.CO},
       adsurl = {https://ui.adsabs.harvard.edu/abs/2025MNRAS.544.2586N}
}

@ARTICLE{Hand2018AJ,
       author = {{Hand}, Nick and {Feng}, Yu and {Beutler}, Florian and {Li}, Yin and {Modi}, Chirag and {Seljak}, Uro{\v{s}} and {Slepian}, Zachary},
        title = "{nbodykit: An Open-source, Massively Parallel Toolkit for Large-scale Structure}",
      journal = {\aj},
         year = 2018,
        month = oct,
       volume = {156},
       number = {4},
          eid = {160},
        pages = {160},
          doi = {10.3847/1538-3881/aadae0},
archivePrefix = {arXiv},
       eprint = {1712.05834},
 primaryClass = {astro-ph.IM},
       adsurl = {https://ui.adsabs.harvard.edu/abs/2018AJ....156..160H}
}

@ARTICLE{2026arXiv260315732B,
       author = {{Bayer}, Adrian E. and {Parker}, Liam and {Valcin}, David and {Chen}, Shi-Fan and {Modi}, Chirag and {Seljak}, Uros},
        title = "{Field-Level Inference from Galaxies: BAO Reconstruction}",
      journal = {arXiv e-prints},
         year = 2026,
        month = mar,
          eid = {arXiv:2603.15732},
        pages = {arXiv:2603.15732},
          doi = {10.48550/arXiv.2603.15732},
archivePrefix = {arXiv},
       eprint = {2603.15732},
 primaryClass = {astro-ph.CO},
       adsurl = {https://ui.adsabs.harvard.edu/abs/2026arXiv260315732B}
}

@ARTICLE{2018JCAP...10..028M,
       author = {{Modi}, Chirag and {Feng}, Yu and {Seljak}, Uro{\v{s}}},
        title = "{Cosmological reconstruction from galaxy light: neural network based light-matter connection}",
      journal = {\jcap},
         year = 2018,
        month = oct,
       volume = {2018},
       number = {10},
          eid = {028},
        pages = {028},
          doi = {10.1088/1475-7516/2018/10/028},
archivePrefix = {arXiv},
       eprint = {1805.02247},
 primaryClass = {astro-ph.CO},
       adsurl = {https://ui.adsabs.harvard.edu/abs/2018JCAP...10..028M}
}

@article{10.1093/mnras/stt449,
    author = {Jasche, Jens and Wandelt, Benjamin D.},
    title = {Bayesian physical reconstruction of initial conditions from large-scale structure surveys},
    journal = {Monthly Notices of the Royal Astronomical Society},
    volume = {432},
    number = {2},
    pages = {894-913},
    year = {2013},
    month = {06},
    issn = {0035-8711},
    doi = {10.1093/mnras/stt449},
    url = {https://doi.org/10.1093/mnras/stt449},
    eprint = {https://academic.oup.com/mnras/article-pdf/432/2/894/18450603/stt449.pdf},
}

@ARTICLE{2021arXiv210412864M,
       author = {{Modi}, Chirag and {Lanusse}, Fran{\c{c}}ois and {Seljak}, Uro{\v{s}} and {Spergel}, David N. and {Perreault-Levasseur}, Laurence},
        title = "{CosmicRIM : Reconstructing Early Universe by Combining Differentiable Simulations with Recurrent Inference Machines}",
      journal = {arXiv e-prints},
         year = 2021,
        month = apr,
          eid = {arXiv:2104.12864},
        pages = {arXiv:2104.12864},
          doi = {10.48550/arXiv.2104.12864},
archivePrefix = {arXiv},
       eprint = {2104.12864},
 primaryClass = {astro-ph.CO},
       adsurl = {https://ui.adsabs.harvard.edu/abs/2021arXiv210412864M}
}








\bsp	
\label{lastpage}
\end{document}